\documentclass{kapproc} 

\usepackage{psfig}

\usepackage[dvips]{graphicx}

\upperandlowercase

\long\def\Ignore#1{\relax}
\def\etal{{\it et al.}}
\def\eg{{\it e.g.}}
\def\etc{{\it etc.}}
\def\ie{{\it i.e.}}

\kluwerbib 

\begin{document}

\articletitle{Anomalously Weak Dynamical \hfil\break
Friction in Halos}


\author{J. A. Sellwood}
\affil{Department of Physics \& Astronomy, Rutgers University\\
136 Frelinghuysen Road, Piscataway NJ 08854-8019, USA}
\email{sellwood@physics.rutgers.edu}

\author{Victor P. Debattista}
\affil{Astronomy Department, University of Washington\\
Box 351580, Seattle, WA 98195-1580, USA}
\email{debattis@astro.washington.edu}

\begin{abstract}
A bar rotating in a pressure-supported halo generally loses angular
momentum and slows down due to dynamical friction.  Valenzuela \&
Klypin report a counter-example of a bar that rotates in a dense halo
with little friction for several Gyr, and argue that their result
invalidates the claim by Debattista \& Sellwood that fast bars in real
galaxies require a low halo density.  We show that it is possible for
friction to cease for a while should the pattern speed of the bar
fluctuate upward.  The reduced friction is due to an anomalous
gradient in the phase-space density of particles at the principal
resonance created by the earlier evolution.  The result obtained by
Valenzuela \& Klypin is probably an artifact of their adaptive mesh
refinement method, but anyway could not persist in a real galaxy.  The
conclusion by Debattista \& Sellwood still stands.
\end{abstract}


\section{Introduction}
It is now well established that a bar rotating in a halo loses angular
momentum through dynamical friction.  This topic has received a lot of
attention recently for two important reasons: (1) it offers a
constraint on the density of the DM halo (Debattista \& Sellwood 1998,
2000), and (2) it may flatten the density cusp (Weinberg \& Katz
2002).

Both these claims have been challenged.  Realistic bars in cuspy halos
produce a mild density decrease at most (Holley-Bockelmann \etal\
2003) or even a slight increase (Sellwood 2003), but we leave this
issue aside here and concentrate instead on the density constraint.
Holley-Bockelmann \& Weinberg (2005) announce a preliminary report of
simulations with weak friction in halos having uniform density
cores, but we focus here on the older counter-example claimed by
Valenzuela \& Klypin (2003; hereafter VK03) of a bar that experiences
little friction in a cusped dense halo.

\begin{figure}[t]
\centerline{\psfig{figure=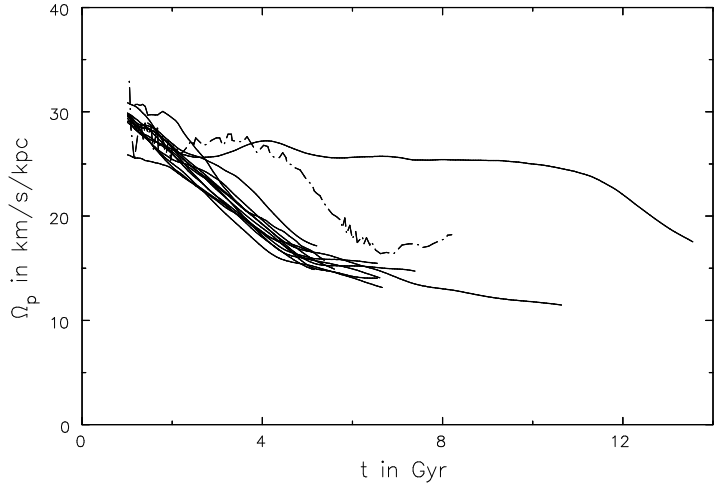,width=.8\hsize,clip=}}
\caption{The time evolution of the bar pattern speed in a number of
resimulations of model A$_1$ of VK03.  The evolution reported by VK03
is reproduced as the dot-dashed line; all other lines are from
simulations from the same initial particle load, but run with our code
using many different sets of numerical parameters.}
\label{VKcomp}
\end{figure}

VK03 kindly made available the initial positions and velocities of all
the particles of their model A$_1$, in which the bar did not slow for
2-3 Gyrs after it had formed and settled.   We have used our code
(Sellwood 2003) to rerun this simulation many times, and the pattern
speed evolution in many of these runs is shown in Figure~\ref{VKcomp}.
It is striking that in most cases, the bar slowed earlier than VK03
found, but in one anomalous case, the bar stayed fast for about 10
Gyr!  The anomalous result is not a consequence of some inadequate
numerical parameter, since many of the other cases are from models
with parameters that bracket those of the anomalous case -- \ie\
longer and shorter time steps, coarser and finer grids, \etc

Note that apart from the crucial delay in the onset of friction in the
case by VK03 and the one anomalous case we find, the evolution is
generally very similar.  In particular, whenever the bar slows, it
slows at a similar rate.  The following sections account for the
discrepancies between the results shown in Fig.~\ref{VKcomp}.

\section{Frictional Torque}
In a classic paper, Tremaine \& Weinberg (1984) laid out the
mathematical apparatus for friction in a spherical system.  Following
the precepts of Lynden-Bell \& Kalnajs (1972), they derived a formula
for the torque experienced by a rotating perturbation potential,
$\Phi_p$.  They work in action-angle variables (see Binney \& Tremaine
1987, \S3.5).  In a spherical potential, there are two non-zero
actions: the total angular momentum per unit mass $L \equiv J_\phi$
and the radial action $J_r$, each associated with two separate
frequencies, $\Omega$ and $\kappa$, which are generalizations to
orbits of arbitrary eccentricity of the usual frequencies of Lindblad
epicycles familiar from disk dynamics.  In the limit that a constant
amplitude perturbation rotates steadily at $\Omega_p$, they showed
that the net LBK torque is
\begin{equation}
\tau_{\rm LBK} \propto \sum_{m,k,n} \left(m{\partial f \over \partial
L} + k {\partial f \over \partial J_r}\right)|\Phi_{mnk}|^2 \delta(n
\Omega + k \kappa - m \Omega_p),
\end{equation}
where $f$ is the usual distribution function and $\Phi_{mnk}$ is a
Fourier coefficient of the perturbing potential.  The Dirac delta
function implies that the net torque is the sum of the separate
contributions from resonances, where $n \Omega + k \kappa = m
\Omega_p$.  Because the bar pattern speed decreases, as a result of
the frictional torque, this expression needs to be generalized to a
time-dependent forcing (see Weinberg 2004), but the revised expression
for the torque still contains the same derivatives of the distribution
function.

Lynden-Bell (1979) offered a clear insight into how an orbit is
affected when close to a resonance.  The unperturbed orbit, which is a
rosette in an inertial reference frame, closes in any frame that
rotates at the rate
\begin{equation}
\Omega^\prime = \Omega + k \kappa / m,
\end{equation}
for any pair $k,\,m$.  [See \eg\ Kalnajs (1977) for illustrations of
several of the most important shapes.]  When the pattern speed of the
bar is close to $\Omega^\prime$ for some pair $k,\,m$, the orbit can
be regarded as a closed figure that precesses at the slow rate
\begin{equation}
\Omega_s \equiv (\Omega^\prime - \Omega_p) \ll \Omega_p.
\end{equation}
Under these circumstances, the ``fast action'' is adiabatically
invariant, while the ``slow action'' can suffer a large change.
Things are particularly straightforward at corotation, where the fast
action is the radial action, while the slow action that can suffer a
large change is simply the angular momentum.

\begin{figure}[t]
\centerline{\psfig{figure=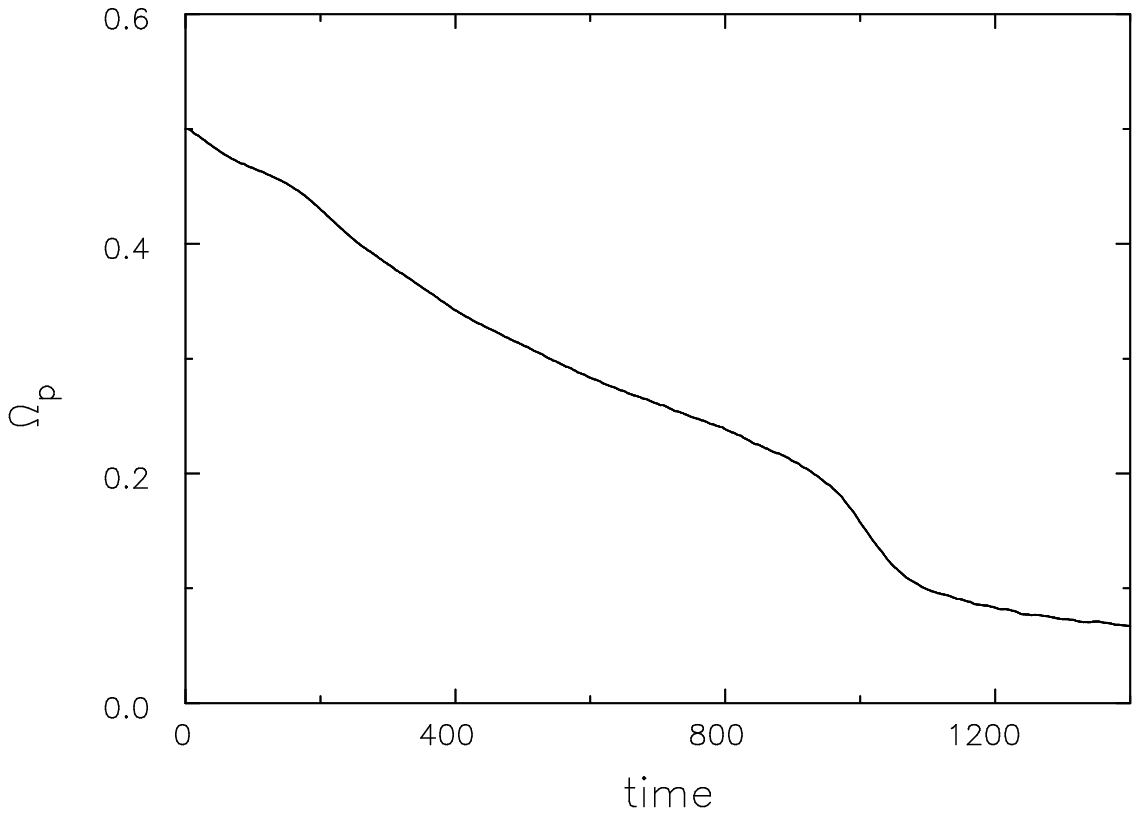,width=.5\hsize,clip=} \hfil
\psfig{figure=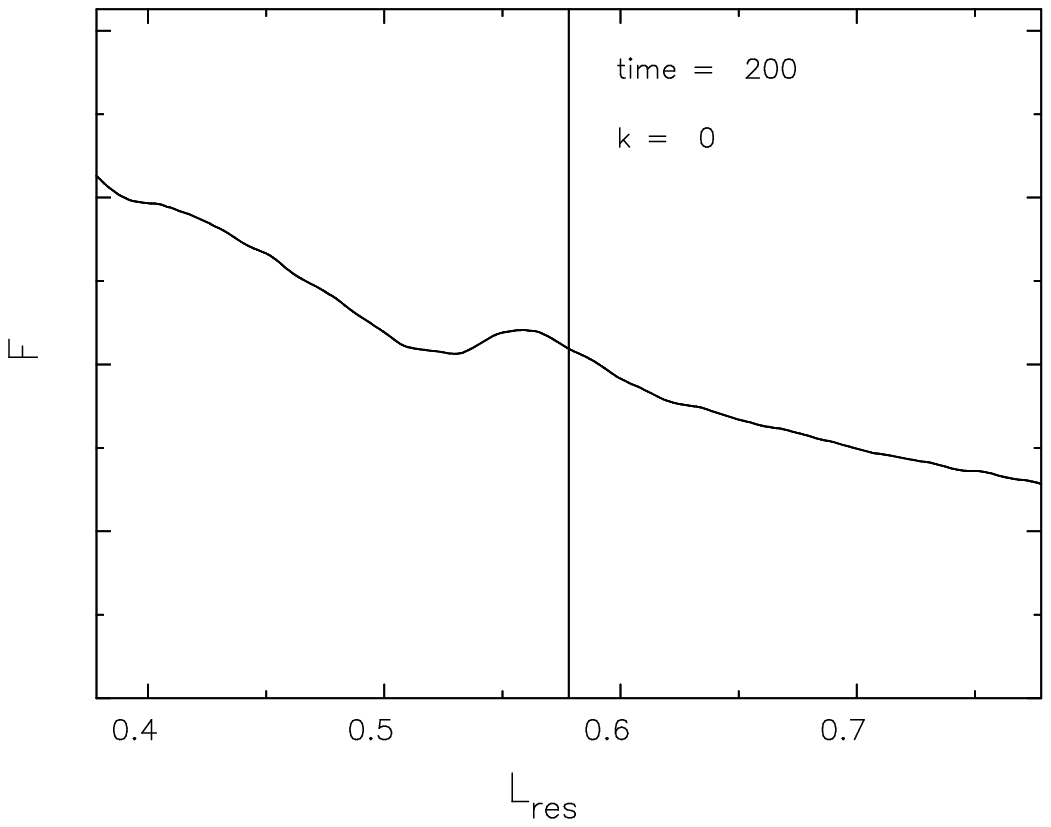,width=.5\hsize,clip=}}
\caption{Left: The time evolution of the bar pattern speed in the
restricted simulation discussed in \S3.  This simulation employs 10M
particles.  Right: The mean density of particles as a function of
$L_{\rm res}$ at $t=200$ in the same simulation.}
\label{restrict}
\end{figure}

\section{Restricted Simulations}
Fully self-consistent simulations are complicated by evolution of the
total potential, changes to the bar mass profile, \etc\ ~It is
therefore easier first to try to understand ``restricted'' simulations
in which a rigid bar rotates in halo of non-interacting test particles
(Lin \& Tremaine 1983; Sellwood 2004).  The particles move in a rigid
halo potential that is perturbed by that of the rotating bar, and the
bar is accelerated in response to the vector-sum of the
non-axisymmetric forces felt by the particles.  Figure~\ref{restrict}
shows an example of the pattern speed evolution of a homogeneous
ellipsoid with principal axes $1:0.5:0.1$ in a Hernquist (1990) halo.
The bar mass is 1\% of the halo mass, $M_h$, has a semi-major axis
equal to the halo break radius $r_h$, and rotates initially so that it
just fills its corotation circle.  We use units such that
$G=M_h=r_h=1$.

At intervals during the simulation, we compute $\Omega^\prime = \Omega
+ k \kappa / m$ for every particle, and calculate $F$, the average
density of particles as a function $\Omega^\prime$.  It is somewhat
easier to understand a plot of $F$ as a function of the angular
momentum, $L_{\rm res}$, of a circular orbit that has the given
$\Omega^\prime(k,m)$.

The right-hand panel of Figure~\ref{restrict} shows the form of $F$
near to corotation ($m = 2, \; k = 0$) at $t=200$, which is typical.
Near to the resonance (marked by the vertical line), particles cross
corotation in both directions on horse-shoe orbits (Binney \& Tremaine
1987, \S7.5).  The generally negative slope of $F$ implies an excess
of lower $L$ particles that gain angular momentum and move out across
the resonance, and this imbalance is responsible for friction.  If the
pattern speed were to stay constant, the imbalance would tend to
flatten the slope of $F$, and the distribution of particles about the
resonance would approach kinetic equilibrium in which there would be
more nearly equal numbers of particles crossing in both directions.
But as $\Omega_p$ declines, the resonance keeps moving to larger
$L_{\rm res}$, and equilibrium is never established; instead, the
density of particles about the dominant resonance(s) responsible for
friction takes on the characteristic humped form shown in
Fig.~\ref{restrict}.

Friction arises principally at corotation over most of the evolution.
The outer Lindblad resonance contributes in the early stages, but
dominates only if the bar is unreasonably fast.  The inner Lindblad
resonance becomes important only when the bar is already quite slow;
\eg\ it is responsible for the more rapid braking around $t=1000$ in
Fig.~\ref{restrict}.

\begin{figure}[t]
\centerline{\psfig{figure=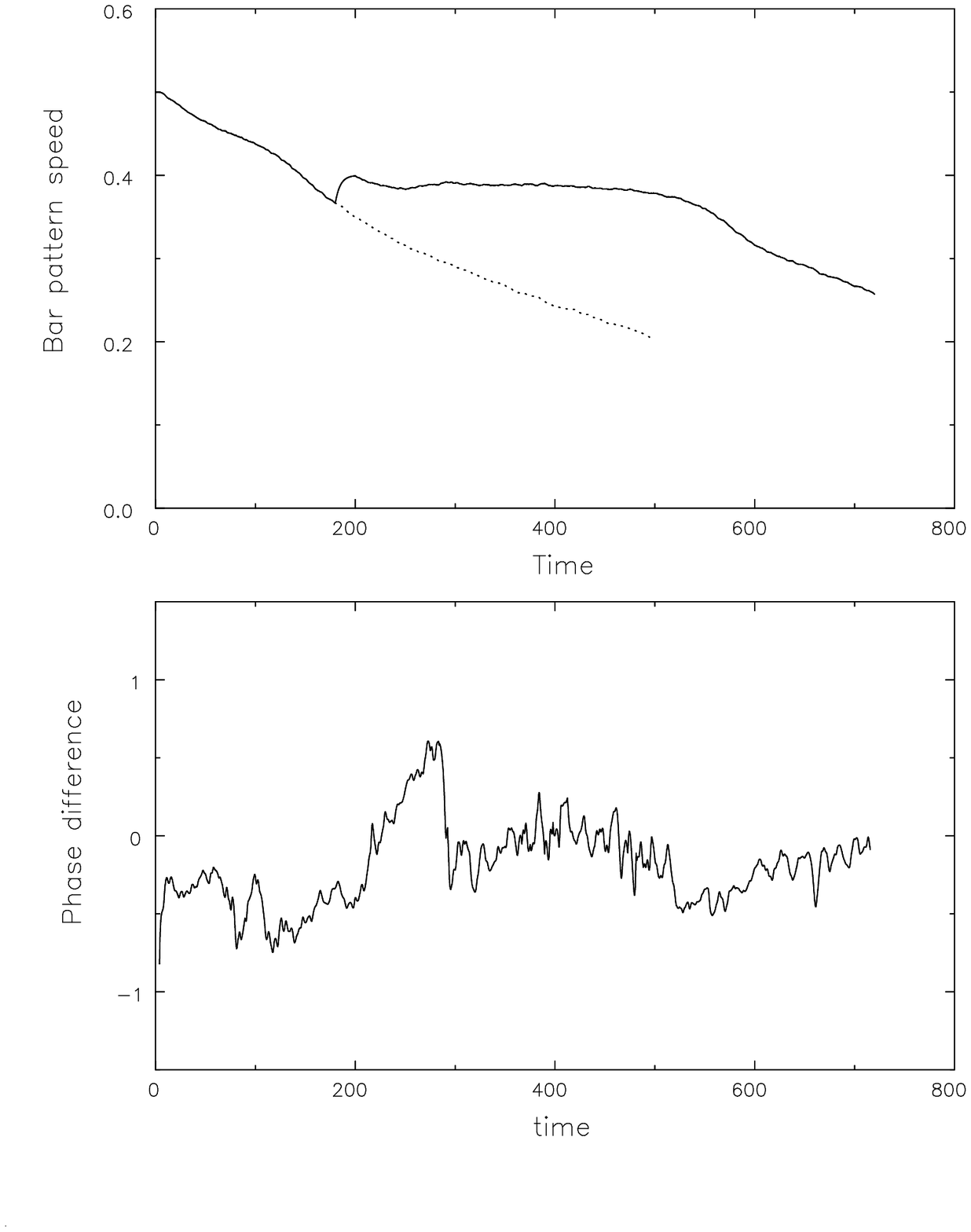,width=.8\hsize,clip=}}
\caption{The solid curve shows evolution of the bar pattern speed in a
restricted experiment in which the bar was reaccelerated (by external
interference) between times 180 and 200, but was otherwise allowed to
evolve freely.  (This experiment used 1M particles.) The dotted curve
shows what happens without interference.}
\label{drive}
\end{figure}

\section{Anomalous Situation}
\begin{figure}[t]
\centerline{\psfig{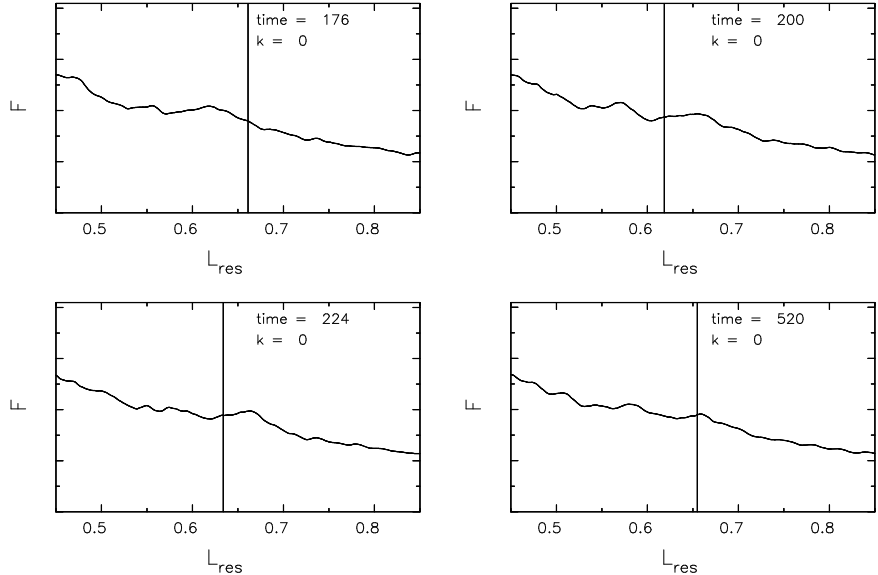}}
\caption{The mean density of particles as a function of $L_{\rm res}$
at several different times in the simulation shown in Fig.~\ref{drive}.}
\label{Ls_forced}
\end{figure}

Now suppose that $\Omega_p$ rises for some reason, after having
declined for some time, as illustrated in Figure~\ref{drive}.  The
shoulder in $F$ created by the previous friction survives, but the
resonance at the now higher $\Omega_p$ lies on the other side of the
shoulder, as shown in Figure~\ref{Ls_forced}.  Thus the local gradient in
$F$ has changed sign, leading to an adverse gradient for friction, and
the torque from the dominant resonance disappears.  Under these
circumstances, a balance between gainers and losers is soon
established, and the bar can rotate in a dense halo with little
friction, which we describe as a ``metastable state''.

In fact, $\Omega_p$ declines slowly because of weak friction at other
resonances, and normal friction resumes when the slope of $F$ at the
main resonance changes, as shown in the last frame of
Fig.~\ref{Ls_forced}.

\section{Self-consistent Simulations}
If we now re-examine Fig.~\ref{VKcomp}, we see that the period of weak
friction is preceded by a small rise in the bar pattern speed in both
the simulation of VK03 (dot-dashed line) and in the anomalous case we
found.  It is likely therefore that friction stopped for
a while in both cases because the local density gradient across the
principal resonance became flat, as just described.

Analysis of our simulation that displayed this behavior suggests that
$\Omega_p$ rose because of an interaction between the bar and a spiral
in the disk, which caused the bar to gain angular momentum.  Such an
event is rare; spirals generally remove angular momentum from the bar
at most relative phases.  It is possible that VK03 were unlucky to
have such an event in their case, but they report similar behavior in
their model B making a chance event unlikely.

One significant difference between our code and that used by VK03
(Kravtsov, Klypin \& Khokhlov 1997) is that their resolution is
adaptive, which causes gravity to strengthen at short range when the
grid is refined.  The increase in the local density as the bar
amplitude rises causes the code to refine the grid, strengthening
gravity and thereby causing the bar to contract slightly and to
spin-up.  We have found that a reduction of softening length in our
code at this epoch also leads to a metastable state.  It is likely,
therefore, that their anomalous result is an artifact of their
adaptive code.

\section{The Metastable State is Fragile}
Whatever the origin of the bar speed-up in simulations, it remains
possible that the metastable state could occur in real galaxies.  If
friction in a dense halo can be avoided for this reason, then the
observed pattern speeds will provide no constraint on the halo
density.

However, further experiments in which we perturbed our model in the
metastable state very slightly, revealed that the state is highly
fragile.  For example, a satellite of merely 1\% of the galaxy mass
flying by at 30kpc is sufficient to jolt the system out of the
metastable state.  We therefore conclude that anomalously weak
friction is unlikely to persist for long in nature.

\section{Conclusions}
Tremaine \& Weinberg (1984) showed that angular momentum is
transferred from a rotating bar to the halo through resonant
interactions.  We find that friction is dominated by a single
resonance at most times, and that corotation is most important for a
bar with realistic pattern speed -- \ie\ when the bar extends almost
to corotation.

Friction arises because the phase space density is a decreasing
function of angular momentum in normal circumstances, causing an
excess of particles that gain angular momentum over those that lose.
While this process would tend to flatten the density gradient if the
pattern speed remained steady, the decreasing angular speed of the bar
prevents this steady state from being reached.  Instead we find that
the density of particles in phase space develops a shoulder, with the
resonance holding station on the high-angular momentum side of the
shoulder as the feature moves to larger $L_{\rm res}$.

However, if the bar is spun up slightly for some reason after a period
of normal friction, the rise in the pattern speed may move the
resonance to the other side of the pre-constructed shoulder.  The
change in the local gradient of particle density at the dominant
resonance causes friction to become very weak for a while, allowing
the bar to rotate almost steadily.  Mild friction persists because of
contributions from other, sub-dominant resonances, and normal friction
resumes once the pattern speed has declined sufficiently for the
gradient at the main resonance to become favorable for friction once
more.

A state in which strong friction is suspended for this reason is
``meta\-stable'', both because it relies on a local minimum in the phase
space density, and because the state is fragile.  A very mild jolt to
the system is sufficient to cause normal friction to resume.

The absence of friction in the simulation A$_1$ reported by Valenzuela
\& Klypin (2003) is probably an artifact of their code.  Their
adaptive grid causes gravity to strengthen as the bar density builds
up, making the pattern speed of the bar rise for a purely numerical
reason.  Thus their claimed counter-example to the argument of
Debattista \& Sellwood (1998, 2000) is a numerical artifact of their
method.  {\it Pace} Holley-Bockelmann \& Weinberg (2005), our
constraint on halo density still stands: A {\it strong\/} bar in a
{\it dense\/} halo will quickly become unacceptably slow through
dynamical friction.

\begin{acknowledgments}
We thank Anatoly Klypin for providing the initial positions and
velocities of the particles in his model, and for many discussions.
This work was supported by NASA (NAG 5-10110) and the NSF
(AST-0098282).
\end{acknowledgments}


\begin{chapthebibliography}{1}
\def\apj{{\it Ap.\ J.}}
\def\apjl{{\it Ap.\ J. Lett.}}
\def\apjs{{\it Ap.\ J. Suppl.}}
\def\mnras{{\it MNRAS}}

\bibitem{BT}
Binney, J. \& Tremaine, S. 1987, {\it Galactic Dynamics\/} (Princeton: Princeton University Press)

\bibitem{DS98}
Debattista, V. P. \& Sellwood, J. A. 1998, \apjl, {\bf 493}, L5

\bibitem{DS00}
Debattista, V. P. \& Sellwood, J. A. 2000, \apj, {\bf 543}, 704

\bibitem{H90}
Hernquist, L. 1990, \apj, {\bf 356}, 359

\bibitem{HBWK}
Holley-Bockelmann, K., Weinberg, M. D. \& Katz, N. 2003, astro-ph/0306374

\bibitem{HBW}
Holley-Bockelmann, K. \& Weinberg, M. D. 2005, DDA abstract 36.0512

\bibitem{K77}
Kalnajs, A. J. 1977, \apj, {\bf 212}, 637

\bibitem{KKK97}
Kravtsov, A. V., Klypin, A. \& Khokhlov, A. M. 1997, \apjs, {\bf 111}, 73

\bibitem{LT83}
Lin, D. N. C. \& Tremaine, S. 1983, \apj, {\bf 264}, 364

\bibitem{LB79}
Lynden-Bell, D. 1979, \mnras, {\bf 187}, 101

\bibitem{LBK}
Lynden-Bell, D. \& Kalnajs, A. J. 1972, \mnras, {\bf 157}, 1

\bibitem{S03}
Sellwood, J. A. 2003, \apj, {\bf 587}, 638

\bibitem{S04}
Sellwood, J. A. 2004, astro-ph/0407533

\bibitem{TW84}
Tremaine, S. \& Weinberg, M. D. 1984, \mnras, {\bf 209}, 729

\bibitem{VK03}
Valenzuela, O. \& Klypin, A. 2003, \mnras, {\bf 345}, 406

\bibitem{W04}
Weinberg, M. D. 2004, astro-ph/0404169

\bibitem{WK03}
Weinberg, M. D. \& Katz, N. 2002, \apj, {\bf 580}, 627

\end{chapthebibliography}
\end{document}